\documentclass{article}
\usepackage{spconf,amsmath,graphicx}
\usepackage{amsmath,graphicx}
\usepackage{amsthm}
\usepackage{cite}
\usepackage{graphicx}
\usepackage{amssymb}
\usepackage{amsfonts}
\usepackage{multirow}
\usepackage{mathrsfs}
\usepackage{color}
\usepackage{algorithm}
\usepackage{algorithmic}
\usepackage{multirow}
\usepackage{amsmath}
\usepackage{graphics}
\usepackage{graphicx}
\usepackage{epsfig}
\usepackage{cases}
\usepackage{amsmath,bm}
\usepackage{wasysym}
\usepackage{graphicx}
\usepackage{subfigure}
\usepackage{hyperref}
\hypersetup{ hidelinks = true, }

\newtheorem{mytheorem}{Theorem}

\title{Distributed Convergence Verification for Gaussian Belief Propagation}

\name{Jian Du,
 Soummya Kar and  Jos{\'e} M. F. Moura }
\address{Electrical and Computer Engineering, Carnegie Mellon University$^{\dagger}$, Pittsburgh, PA}
\begin{document}
\newcommand*{\QEDA}{\hfill\ensuremath{\blacksquare}}
\def\N{{\mathcal{N}}}
\def\B{{\mathcal{B}}}
\def\I{{\textbf{I}}}
\def\diag{{\textrm{diag}}}
\def\i {{ -i}}

%
\maketitle
\begin{abstract}
Gaussian belief propagation (BP) is a computationally efficient method to approximate  the marginal distribution and has been widely used for inference with high dimensional data as well as  distributed estimation in large-scale networks. 
However, the convergence of Gaussian BP is still an open issue. 
Though sufficient convergence conditions have been studied in the literature,
verifying these conditions requires gathering all the information over the 
whole network, which defeats the main advantage of distributed computing by using Gaussian BP.
In this paper, we propose a novel sufficient convergence condition for Gaussian BP that applies to  both the pairwise linear Gaussian model and to  Gaussian Markov random fields. We show analytically that this sufficient convergence condition can be easily verified in a distributed way that satisfies the network topology constraint.
\end{abstract}

\section{Introduction}\label{Section 1}
Gaussian belief propagation (BP) \cite{journalversion, WalkSum1} provides an efficiently distributed way to compute the marginal distribution of  unknown variables with a joint Gaussian distribution, and Gaussian BP has been adopted in a variety of topics 
including image denosing \cite{lan2006efficient}, distributed power state estimation \cite{A1} in smart grid, data detection for massive MIMO systems\cite{som2011low}, distributed beamforming \cite{A2} and synchronization  \cite{JianClock,cfo, clockICASSP,du2017proactive} in distributed networks,  fast solver for system of linear equations \cite{A4}, distributed rate control in ad-hoc networks \cite{A5},  factor analyzer networks \cite{A6}, sparse Bayesian learning \cite{A7},  inter-cell interference mitigation \cite{A8},
and peer-to-peer rating in social networks \cite{A9}.
It has been shown that Gaussian BP computes the optimal centralized estimator if it converges \cite{journalversion, WalkSum1}.
The challenge of deploying Gaussian BP for large-scale networks is determining if it will converge. In particular, it is generally known that, if the factor graph, which represents the joint distribution, contains cycles, the Gaussian BP algorithm may diverge. Thus, determining convergence conditions for the Gaussian BP algorithm is very important. Further, in practical applications for large-scale networks, having an easily verified sufficient convergence condition is of great importance.

Sufficient convergence conditions for Gaussian BP have been developed in \cite{DiagnalDominant,WalkSum1,minsum09}  when the underlying Gaussian distribution is expressed in terms of Gaussian Markov random fields (GMRFs) with    \textit{scalar} variables.
These works focus on the convergence analysis of Gaussian BP for computing the marginal distribution of a joint distribution with pairwise factors.
However, as demonstrated in \cite{journalversion}, the iterative equations for Gaussian BP on GMRFs are different from those for distributed estimation problems such as in \cite{A1,A2,JianClock,cfo,A4,A5,A6,A7,A8},  where  linear Gaussian measurements  are involved.

Recently, the convergence properties of Gaussian BP for the linear Gaussian measurement model with \textit{vector variables} have been  studied in  \cite{journalversion}.
It is analytically shown that the marginal covariance for each unknown vector variable converges to a unique positive definite matrix with double exponential rate with arbitrary positive-semidefinite initial message information matrix.
Further, a sufficient  convergence condition that guarantees  the belief mean to converge to the exact marginal mean is also given. 
It is also shown that this sufficient convergence condition subsumes all the previous sufficient convergence conditions in \cite{DiagnalDominant,WalkSum1,minsum09}.

Though  sufficient convergence conditions have been studied for both the GMRFs and linear Gaussian models,  verifying these conditions all require to gather all the information together to perform centralized computation. For example,  in \cite{journalversion}, the convergence condition requires computation with complexity of $\mathcal O(N^3)$  of the spectrum radius of a dimension-$N$ matrix, where $N$ is the dimension of a vector constructed by stacking together all the mean vectors exchanged in the network at each iteration. This defeats the main advantage of low computation complexity of a distributed inference algorithm. In this paper, we present a sufficient convergence condition for Gaussian BP on both GMRFs and linear Gaussian models and demonstrate that it can be easily verified in a distributed way that respects the network topology constraints.

\section{Computation Model }\label{hybrid}
\subsection{Belief Propagation}
The  BP algorithm computes the marginal distribution/function given a factorization of the distribution/function. 
The factorization can be represented by a graphical model known as factor graph
\cite{Kschischang},
and the BP algorithm can be expressed as message passing on the factor graph.

In the factor graph,  every vector variable  $\textbf{x}_i$ is represented by a circle (called variable node)
and the probability distribution of a vector variable or a group of vector variables is represented by a square (called factor node).
A variable node is connected to a factor node if the variable is involved in that particular factor.
For example, Fig.~1 shows the factor graph representation of a joint Gaussian distribution.
The explicit functions  of $f_{i,j}$ and $f_i$ for the linear Gaussian model and the GMRF will be given later.

We next derive the 
BP algorithm   over the corresponding factor graph. It
involves two types of messages:
one is the message
from a variable node $\textbf x_j$ to its neighboring factor node $f_{i,j}$, defined as
\begin{equation} \label{BPv2f1}
m^{\left(\ell\right)}_{j \to f_{i,j}}\left(\textbf x_j\right)
= p\left(\textbf x_j\right)
\prod_{f_{k,j}\in \mathcal B(j)\setminus f_{i,j}}m^{\left(\ell-1\right)}_{f_{k,j}\to j}\left(\textbf x_j\right),
\end{equation}
where $\mathcal B\left(j\right)$ denotes the set of neighbouring factor nodes  of $\textbf x_j$,
and $m^{\left(\ell-1\right)}_{f_{k,j}\to j}\left(\textbf x_j\right)$ is the   message  from $f_{k,j}$ to $\textbf x_j$ at iteration $\ell-1$.
The second type of
message is from a factor node $f_{i,j}$ to a neighboring variable node $\textbf{x}_i$, defined as
\begin{equation}\label{BPf2v1}  
m^{\left(\ell\right)}_{f_{i,j} \to i}\left(\textbf{x}_i\right)
=  \int \!\!\!\cdots\!\!\! \int
f_{i,j} \times \!\!\!\!\!
\prod_{j\in\mathcal B\left(f_{i,j}\right)\setminus i}\!\!\!\!\!\!\! m^{ \left(\ell\right)}_{j \to f_{i,j}}\left(\textbf x_j\right)
\,\!\!\mathrm{d}\left\{\textbf x_j\right\}_{j\in\B\left(f_{i,j}\right)\setminus i},
\end{equation}
where $\mathcal B\left(f_{i,j}\right)$ denotes the set of neighboring variable nodes of $f_{i,j}$.
The process iterates between equations (\ref{BPv2f1}) and (\ref{BPf2v1}).
At each iteration $\ell$, the approximate marginal distribution, also referred to as belief, on $\textbf{x}_i$ is computed locally at $\textbf{x}_i$ as
\begin{equation} \label{BPbelief}
b_{\textrm{BP}}^{\left(\ell\right)}\left(\textbf{x}_i\right)
= p\left(\textbf{x}_i\right) \prod_{ f_n\in \mathcal B\left(i\right)} m^{\left(\ell\right)}_{ f_n \to i}\left(\textbf{x}_i\right).
\end{equation}
In the following subsections, we show the updating equations of Gaussian BP for the linear Gaussian model and the GMRF.

\subsection{ Linear Gaussian Model}
Consider a general connected network
of $M$  nodes, with $\mathcal{V}=\{1,\ldots, M\}$ denoting the set of nodes,
and $\mathcal{E}_{\textrm{Net}} \subset \mathcal{V} \times  \mathcal{V}$  the set of all undirect communication links in the network, i.e., if $i$ and $j$ are within the communication range, $(i, j) \in \mathcal{E}_{\textrm{Net}}$.
The  local observations, $\textbf{y}_{i,j}$, between node $i$ and $j$ are modeled by the pairwise  linear Gaussian model \cite{globalsip,powerjian}:
\begin{equation} \label{linear}
\textbf{y}_{i,j} =
\textbf{A}_{j,i}\textbf{x}_i
+\textbf{A}_{i,j}\textbf{x}_j
+ \textbf{z}_{i,j},
\end{equation}
where
$\textbf{A}_{j,i}$ and $\textbf{A}_{i,j}$ are  known coefficient matrices with full column rank,
$\textbf{x}_i$ and $\textbf{x}_{j}$ are  local unknown vector parameters at node $i$ and $j$ with  dimension $N_i \times 1$ and $N_j \times 1$, and with  prior distribution $p(\textbf{x}_i)\sim \mathcal{N}(\textbf{x}_i|\textbf{0},\textbf{W}_{i})$ and
$p(\textbf{x}_j)\sim \mathcal{N}(\textbf{x}_j|\textbf{0},\textbf{W}_{j})$,
and $\textbf{z}_{i,j}$ is  additive noise with distribution $\textbf{z}_{i,j}\sim \mathcal{N}(\textbf{z}_{i,j}|\textbf{0},\textbf{R}_{i,j})$.
It is assumed that
$p(\textbf{x}_i, \textbf{x}_j)=p(\textbf{x}_i)p(\textbf{x}_j)$
and
$p(\textbf{z}_{i,j},\textbf{z}_{s,t})
=p(\textbf{z}_{i,j})p(\textbf{z}_{s,t})$ for $\{i,j\}\neq \{s,t\}$.
The goal is to estimate $\textbf{x}_i$, based on $\textbf{y}_{i,j}$, $p(\textbf{x}_i)$ and $p(\textbf{z}_{i,j})$ for all $\textbf x_i\in \mathcal V$.
Note that in (\ref{linear}), $\textbf{y}_{i,j}=\textbf{y}_{j,i}$.

The joint distribution  $p\left(\textbf x\right)p\left(\textbf{y}|\textbf x\right)$ is first written as the product of the  prior distribution and the likelihood function as
\begin{equation}\label{jointpost}
p\left(\textbf x\right)p\left(\textbf{y}|\textbf x\right) =
\prod_{i\in \mathcal{V}}
\underbrace{p\left(\textbf x_i\right)}_{\triangleq g_{i}}
\prod_{i\in \mathcal{V}}
\underbrace{p(\textbf{y}_{i,j}| \textbf{x}_i, \textbf{x}_j, \{i,j\}\in  \mathcal{E}_{\textrm{Net}})}_{\triangleq g_{i,j}}.
\end{equation}

\begin{figure}[t]
	\centering
	{\epsfig{file=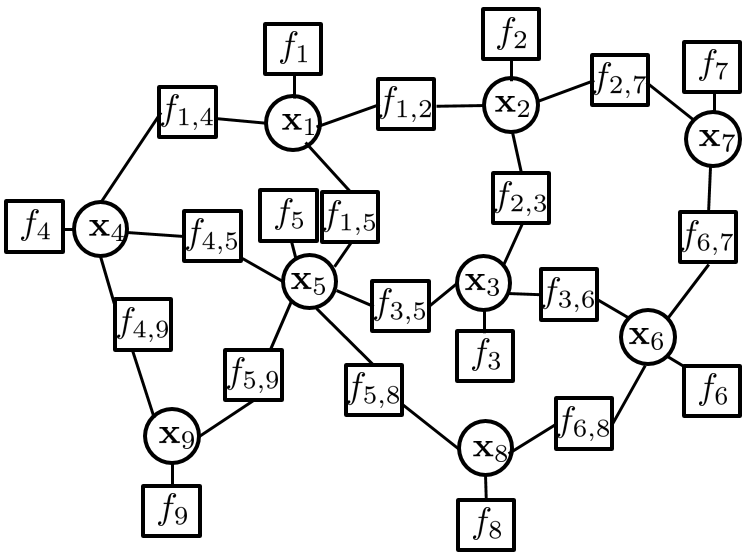, width=2.6in}}
	\caption{An example of factor graph  representing a joint Gaussian distribution.}
	\label{venn}
\end{figure}
Following the computation rule of  (\ref{BPf2v1}),
it can be easily shown  that the
message from  a factor that denotes prior information is a constant, i.e.,
$m^{(\ell)}_{{f}_i \to i}(\textbf{x}_i)
= p(\textbf x_i),\quad \forall \ell=0,1,\ldots\nonumber.$
Further, from (\ref{BPf2v1}) and (\ref{BPv2f1}), message $m^{(\ell)}_{i \to f_i}(\textbf{x}_i)$ will never be used for message updating.
Therefore, we next  focus on the message exchange between the factor denoting the likelihood function and its
neighboring variables.

It can be  shown  that the message  from a factor node to a variable node is given by \cite{journalversion}
\begin{equation} \label{f2v}
m^{(\ell)}_{f_{i,j} \to i}(\textbf{x}_i)\propto
\exp
\big\{-\frac{1}{2}
||\textbf{x}_i- \textbf{v}^{(\ell)}_{f_{i,j}\to i}||
_{\textbf{C}^{(\ell)}_{f_{i,j}\to i}}
\big\},
\end{equation}
where $\textbf{C}_{f_{i,j}\to i}^{(\ell-1)}$ and $ \textbf{v}_{f_{i,j}\to i}^{(\ell-1)}$ are the message covariance matrix and mean vector  received at variable node $i$ at the $\ell-1$ iteration
with
\begin{equation}\label{Cov}
\begin{split}
\left[\textbf{C}^{(\ell)}_{f_{i,j}\to i} \right]^{-1}
=
\textbf{A}_{j,i}^T
\left[ \textbf{R}_{i,j}
+ \textbf{A}_{i,j}\textbf{C}^{(\ell)}_{j\to f_{i,j}}\textbf{A}_{i,j}^T \right]^{-1}
\textbf{A}_{j,i}.
\end{split}
\end{equation}
and
\begin{equation}\label{f2vmm}
\begin{split}
\textbf{v}^{(\ell)}_{f_{i,j}\to i}
=&
\textbf{A}_{j,i}^T
\left[ \textbf{R}_{i,j}
+ \textbf{A}_{i,j}\textbf{C}^{(\ell)}_{j\to f_{i,j}}\textbf{A}_{i,j}^T \right]^{-1}
\\
&\times
\left(\textbf{y}_{i,j}- \textbf{A}_{i,j}
\textbf{v}^{(\ell)}_{j\to f_{i,j}}\right).
\end{split}
\end{equation}
Furthermore, the general expression for the message from a variable node to a factor node  is
\begin{equation} \label{BPvs2f1}
m^{(\ell)}_{j \to f_{i,j}}(\textbf x_j) \propto
\exp
\left\{-\frac{1}{2}
||\textbf x_j- \textbf{v}^{(\ell)}_{j\to f_{i,j}}||
_{\textbf{C}^{(\ell)}_{j\to f_{i,j}}}
\right\},
\end{equation}
where $\textbf{C}_{j\to f_{i,j}}^{(\ell)}$ and $ \textbf{v}_{j\to f_{i,j}}^{(\ell)}$ are the message covariance matrix and mean vector  received at variable node $j$ at the $\ell$-$\textrm{th}$ iteration, where the inverse of the covariance matrix updating equation is given by
\begin{equation} \label{v2fV}
\big[\textbf{C}^{(\ell)}_{j \to f_{i,j}}\big]^{-1}
= \textbf{W}_j^{-1} +
\sum_{f_{k,j}\in\B(j)\setminus f_{i,j}}
\big[\textbf{C}_{f_{k,j}\to j}^{(\ell-1)}\big]^{-1},
\end{equation}
and the mean vector is
\begin{equation}\label{v2fm}
\textbf{v}^{(\ell)}_{j\to f_{i,j}}=
\textbf{C}^{(\ell)}_{j\to f_{i,j}}
\bigg[
\sum_{f_{k,j}\in\B(j)\setminus f_{i,j}}
\big[\textbf{C}_{f_{k,j}\to j}^{(\ell-1)}\big]^{-1}
\textbf{v}^{(\ell-1)}_{f_{k,j}\to j}\bigg].
\end{equation}

Following Lemma 2 in \cite{journalversion}, we know that    setting the initial message covariances inverse (also known as the message information matrix) $[\textbf{C}_{f_{k,j}\to i}^{(0)}]^{-1}\succeq \textbf{0}$  for all $k\in \mathcal{V}$ and $j\in \mathcal B(k)$
guarantees  $[\textbf{C}^{(\ell)}_{j\to f_{i,j}}]^{-1}\succ \textbf{0}$ for $l \geq 1$.
Therefore,
let the initial messages at factor node $f_{k,j}$ be in Gaussian function forms with  covariance $[\textbf{C}_{f_{k,j}\to j}^{(0)}]^{-1} \succeq \textbf{0}$ for all $k \in \mathcal{V}$ and $j \in \mathcal{B}(f_{k,j})$.
Then
all the messages $m^{(\ell)}_{j \to f_{i,j}}(\textbf x_j)$ and
$m^{(\ell)}_{f_{i,j} \to i}(\textbf{x}_i)$ exist and are in Gaussian form.

Furthermore, during each round of message updating, each variable node can compute the belief for $\textbf x_i$
using (\ref{BPbelief}), which can be easily shown to be
\begin{equation}
b_{i}^{(l)}(\textbf x_i)\sim  \mathcal{N}(\textbf x_i|\bm \mu_i^{(l)}, \textbf P_{i}^{(l)}),
\end{equation}
with the
inverse of the covariance matrix
\begin{equation} \label{beliefP}
\left[\textbf P_{i}^{(l)}\right]^{-1} = \sum_{f_{i,j}\in\mathcal B(f_{i,j})}\left[\textbf C_{f_{i,j}\to i}^{(l)}\right]^{-1},
\end{equation}
and mean vector
\begin{equation}\label{beliefu}
\bm \mu_i^{(l)} = \left[\sum_{j\in\mathcal B(f_{i,j})}\big[\textbf C_{f_{i,j}\to i}^{(l)}\big]^{-1}\right]^{-1} \!\!\!\!\!\!\!\sum_{j\in\mathcal B(f_{i,j})}\left[\textbf C_{f_{i,j}\to i}^{(l)}\right]^{-1}\!\!\!\textbf v^{(l)}_{f_{i,j}\to i}.
\end{equation}

The iterative algorithm based on BP is summarized as follows.
The algorithm is started by setting the message from factor node to variable node as
$m^{(0)}_{f_{i,j} \to  i}(\textbf x_i)=\mathcal{N}(\textbf x_i;\bm v^{(0)}_{f_{i,j}\to i}, \bm C_{f_{i,j}\to i}^{(0)})$
with a random initial vector $\bm v^{(0)}_{f_{i,j}\to i}$  and $[\bm C_{f_{i,j}\to i}^{(0)})]^{-1}\succeq \bm 0$.
At each round of message exchange, every variable node computes the outgoing messages to factor nodes
according to (\ref{v2fV}) and (\ref{v2fm}).
After receiving the messages from its neighboring variable nodes, each factor node computes its outgoing messages according to (\ref{Cov}) and (\ref{f2vmm}).
Such iteration is terminated when (\ref{beliefu}) converges (e.g., when $\|\bm\mu_i^{(\ell)}-\bm\mu_i^{(\ell-1)}\|<\eta$, where $\eta$ is a threshold) or the maximum number of iterations is reached.
Then the estimate of $\textbf x_i$ of each node is obtained as in  (\ref{beliefu}).

In  recent work \cite{journalversion, ICASSP}, it is shown that the message information matrix converges with double exponential rate to a unique positive definite matrix with arbitrary positive semidefinite initial value.
Thus, it is reasonable to assume that,
for arbitrary initial value
$\textbf{v}^{\left(0\right)}_{f_{k,j}\to j}$, the evolution of
$\textbf{v}^{\left(\ell\right)}_{j\to f_n}$ in
(\ref{v2fm}) can be written in terms of the limit message information matrices, i.e., $\textbf{C}^{\ast}_{j\to f_{i,j}}$ and $\textbf{C}_{f_{k,j}\to j}^{\ast}$, as
\begin{equation}\label{v2fm1}
\textbf{v}^{(\ell)}_{j\to f_{i,j}}=
\textbf{C}^{\ast}_{j\to f_{i,j}}
\bigg[
\sum_{f_{k,j}\in\B(j)\setminus f_{i,j}}
\big[\textbf{C}_{f_{k,j}\to j}^{\ast}\big]^{-1}
\textbf{v}^{(\ell-1)}_{f_{k,j}\to j}\bigg].
\end{equation}
Using (\ref{f2vmm}), and
replacing indices $j$, $i$ with $k$, $j$ respectively,
$\textbf{v}^{\left(\ell-1\right)}_{f_{k, j}\to j}$  is given by
\begin{equation}\label{f2vmm2}
\begin{split}
\textbf{v}^{(\ell)}_{f_{k,j}\to j}
=&
\textbf{A}_{k,j}^T
\left[ \textbf{R}_{k,j}
+ \textbf{A}_{j,k}\textbf{C}^{\ast}_{k\to f_{k,j}}\textbf{A}_{j,k}^T \right]^{-1}
\\
&\times
\left(\textbf{y}_{k,j}- \textbf{A}_{j,k}
\textbf{v}^{(\ell)}_{k\to f_{k,j}}\right).
\end{split}
\end{equation}
Substituting (\ref{f2vmm2}) into
(\ref{v2fm1}), we have
\begin{equation}\label{v2fm3}
\textbf{v}^{\left(\ell\right)}_{j\to f_{i,j}}=
\textbf{b}_{j \to f_{i,j}}-
\textbf{C}^{\ast}_{j\to f_{i,j}}\!\!\!\!\!\!\!\!
\sum_{f_{k,j}\in\B\left(j\right)\setminus f_{i,j}}\!\!\!\!\!\!
\textbf{C}^{\ast}_{f_{kj}\to j}
\textbf{M}_{k,j}
\textbf{A}_{j,k}
\textbf{v}^{(\ell)}_{k\to f_{k,j}},
\end{equation}
where $\textbf{b}_{j\to f_{i,j}}=
\textbf{C}^{\ast}_{j\to f_{i,j}}
\sum_{f_{k,j}\in\B\left(j\right)\setminus f_{i,j}}
\textbf{M}_{k,j}
\textbf{y}_k $
and
$\textbf M_{k,j}=\textbf{A}_{k,j}^T\left[ \textbf{R}_{k,j}
+ \textbf{A}_{j,k}\textbf{C}^{\ast}_{k\to f_{k,j}}\textbf{A}_{j,k}^T \right]^{-1}$.
The above equation for all $j\in \N(i)$ cases can be further written in compact form as
\begin{equation}\label{v2fm4}
\textbf{v}^{\left(\ell\right)}_{j}=
\textbf{b}_{j} -
\textbf{Q}_{j}
\textbf{v}^{\left(\ell-1\right)},
\end{equation}
with the column vector
$\textbf{v}^{(\ell)}_j$ containing all
$\{\textbf{v}^{(\ell)}_{j\to f_{i,j}}\}_{i\in\N(j)}$
as subvectors
with ascending index   on $i$.
Similarly,
$\textbf{b}_{j}$
containing all
$\{\textbf{b}_{j\to f_{i,j}}\}_{i\in\N(j)}$ as subvectors
with ascending index   on  $i$, and
$\textbf{v}^{\left(\ell-1\right)}$
containing  $\textbf{v}^{\left(\ell-1\right)}_{k\to f_{k,j}}$ for all $f_{k,j}\in\B\left(j\right)\setminus f_{i,j}$ as subvectors
with ascending index first on $z$ and then on $k$.
The matrix   $\textbf{Q}_{j}$
is a row block matrix with component blocks $\textbf 0$ and
$\textbf{C}^{\ast}_{j\to f_{i,j}}$ where
$f_{k,j}\in\B\left(j\right)\setminus f_{i,j}$.

In the next subsection, we   show that the  convergence of Gaussian BP for GMRF also depends on an iterative equation similar to  (\ref{v2fm4}), and a distributed convergence condition for such iterative updating is given in the next section. 
\subsection{Gaussian Markov Random Field}
In the domain of physics and probability, a Markov random field (often abbreviated as MRF), Markov network, or undirected graphical model is a set of random variables having a Markov property described by an undirected graph.
A multivariate normal distribution forms a GMRF with respect to a graph if the missing edges correspond to zeros on the information matrix $\textbf J$.
We denote the normalized joint Gaussian distribution as 
\begin{equation}\label{MRF-eqn}
p\left(\textbf x\right)
\propto
\exp\left(-\frac{1}{2}\textbf x^T\textbf J\textbf x + 
\textbf h^T\textbf x\right)
\end{equation}
where $\textbf J_{i,i} =1$ for all $i$.
Function $p(\textbf x)$ can always be factorized as  
$$
p(\textbf x) \propto 
\prod_{i\in \mathcal{V}} 
{\exp\left(-\frac{1}{2}J_{i,i}x^2_i
	+h_ix_i\right)}
\prod_{\left(i,j\right)\in \mathcal{E}_{\textrm{MRF}} } 
{\exp\left(-x_{i}J_{i,j}x_j\right)}.
$$
We denote the the  factor containing a single variable as
\begin{equation}\label{MRF-eqn}
f_i(x_i)=
{\exp\left(-\frac{1}{2}J_{i,i}x^2_i+h_nx_i\right)}.
\nonumber
\end{equation}
The other factor represents local correlations and contains a pair of unknown variables
\begin{equation}\label{MRF-eqn}
f_{i,j}(x_{i},x_j)={\exp\left(-x_{i}J_{i,j}x_j\right)},
\nonumber
\end{equation}
with $J_{i,j}\neq 0$.
Following the BP message computation rule in (\ref{BPv2f1}) and (\ref{BPf2v1}), we can compute the message updating equations for messages from factor node to variable node $m^{(\ell)}_{f_{k,j} \to j}(\textbf{x}_i)$ and also message from variable node to factor node $m^{(\ell)}_{j \to f_{i,j}}(\textbf x_j)$. 
We omit these equations due to space limitations. 
Similar to the message computation of Gaussian BP for the linear Gaussian model,  the computation of $m^{(\ell)}_{j \to f_{i,j}}(\textbf x_j)$  depends on $m^{(\ell)}_{f_{k,j} \to j}(\textbf{x}_i)$. We substitute $m^{(\ell)}_{f_{k,j} \to j}(\textbf{x}_i)$ into 
$m^{(\ell)}_{j \to f_{i,j}}(\textbf x_j)$ and obtain the message updating expression 
\begin{equation} \label{BPMRF}
m^{(\ell)}_{j \to f_{i,j}}(\textbf x_j) \propto
\exp
\left\{-\frac{1}{2}\Delta J^{(\ell)}_{j\to f_{i,j}}x_j^2 + \Delta h^{(\ell)}_{j\to f_{i,j}}x_j
\right\},
\end{equation}
with the updating parameters
\begin{equation}
\Delta J^{(\ell)}_{j\to f_{i,j}} = -\frac{J_{j,i}^2}{J_{i,i} + \sum_{k\in \mathcal N(i)\setminus j}\Delta J^{(\ell-1)}_{k\to f_{k,i}}},
\end{equation}
and 
\begin{equation}\label{h}
\Delta h^{(\ell)}_{j\to g_{i,j}} = -\frac{J_{j,i}\left(h_i + \sum_{k\in \mathcal N(i)\setminus j} \Delta h^{(\ell-1)}_{k\to g_{k,i}} \right)}{J_{i,i} + \sum_{k\in \mathcal N(i)\setminus j}\Delta J^{(\ell-1)}_{k\to g_{k,i}}}.
\end{equation}

In  \cite[Proposition 1]{WalkSum1},
based on the interpretation that $\left[\textbf J^{-1}\right]_{i,j}$ is the sum of the weights of all the walks from variable $j$ to variable $i$ on the corresponding GMRF, a sufficient Gaussian BP convergence condition  known as walk-summability  is provided, which is  equivalent to
\begin{equation}\label{walksummability}
\textbf I - |{\textbf R}| \succ \textbf 0,
\end{equation}
together with the initial message variance inverse being set to  $0$, where ${\textbf R} = \textbf I - {\textbf J}$ and $|\textbf R|$ is the matrix of entrywise absolute values of $\textbf R$.
However, to verify the walk-summability, one needs to  compute $\textbf I - \textbf J$ in a centralized way, which  defeats the main advantage of distributed computing of Gaussian BP. 

Following the convergence analysis of message information matrix in \cite{journalversion}, it can be shown that $\Delta J^{(\ell)}_{j\to f_{i,j}}$ is convergence guaranteed if the initial value is $\Delta J^{(0)}_{j\to f_{i,j}} = 0$.
Let $\Delta J^{\ast}_{j\to f_{i,j}}$ denote the converged fixed point, then (\ref{h}) can be written as 
\begin{equation}\label{h2}
\Delta h^{(\ell)}_{j\to g_{i,j}} = -\frac{J_{j,i}\left(h_i + \sum_{k\in \mathcal N(i)\setminus j} \Delta h^{(\ell-1)}_{k\to f_{k,i}} \right)}{J_{i,i} + \sum_{k\in \mathcal N(i)\setminus j}\Delta J^{\ast}_{k\to f_{k,i}}}.
\end{equation}
Then similar to the analysis before (\ref{v2fm4}), we have
\begin{equation}\label{v2fm4GMRF}
\{{h}^{(\ell)}_{j\to f_{i,j}}\}_{i\in\N(j)}=
\widetilde{{b}}_{j} -
\widetilde{\textbf{Q}}_{j}
\textbf{h}^{\left(\ell-1\right)},
\end{equation}
with  
$\widetilde{{b}}_{j} = -\frac{J_{j,i}h_i }{J_{i,i} + \sum_{k\in \mathcal N(i)\setminus j}\Delta J^{\ast}_{k\to f_{k,i}}}$, and
$\textbf{h}^{\left(\ell-1\right)}$
containing  ${h}^{\left(\ell-1\right)}_{k\to f_{k,j}}$ for all $f_{k,j}\in\B\left(j\right)\setminus f_{i,j}$ as components.
The matrix   $\widetilde{\textbf{Q}}_{j}$
is a row  matrix with component  $ 0$ and
$-\frac{J_{j,i} \sum_{k\in \mathcal N(i)\setminus j} \Delta h^{(\ell-1)}_{k\to f_{k,i}} }{J_{i,i} + \sum_{k\in \mathcal N(i)\setminus j}\Delta J^{\ast}_{k\to f_{k,i}}}$.

\section{Distributed Convergence Condition}\label{analysis}
The challenge of deploying the Gaussian BP algorithm  is determining whether it will converge.
In particular, it
is generally known that, if the factor graph contains cycles, the BP algorithm may
diverge.
Thus, determining  convergence conditions for the BP algorithm is very important.
In loopy graphs, sufficient conditions for the convergence of Gaussian BP with vector variables   for the linear Gaussian model are available in \cite{journalversion},  and sufficient conditions for the convergence of Gaussian BP with scalar variables  for GMRF are available in \cite{DiagnalDominant, WalkSum1}.
However,  checking these conditions  requires centralized computation, which conflicts with the low computation complexity purpose of distributed computation.
In this section, we  present a sufficient convergence condition for Gaussian BP and demonstrate a distributed way to verify this condition for both the linear Gaussian model and the GMRF model.
In the following, we perform the convergence analysis for the  linear Gaussian model as an example, and all the analysis and results apply to the GMRF case.  

For arbitrary pairwise linear Gaussian model, we  define the  block matrix $\textbf Q$ as
\[\textbf Q =
\begin{bmatrix}
\textbf Q_1\\
\textbf Q_2\\
\vdots\\
\textbf Q_M
\end{bmatrix},\]
and  let $\textbf{v}^{\left(\ell\right)}$ and $\textbf{b}$ be the vector  containing $\textbf{v}^{\left(\ell\right)}_{j }$ and
$\textbf{b}_{j }$ respectively with the same stacking order as $\textbf{Q}_{j} $ in $\textbf Q$.
Following (\ref{v2fm4}), we have
\begin{equation}\label{meanvectorupdate}
\textbf{v}^{\left(\ell\right)} =
-\textbf{Q} \textbf{v}^{\left(\ell-1\right)} + \textbf{b}.
\end{equation}
For this linear updating equation,
it is  known that, for arbitrary initial value $\textbf{v}^{\left(0\right)}$, $\textbf{v}^{\left(\ell\right)}$ converges
if and only if the spectral radius $\rho\left(\textbf{Q}\right)<1$.
In \cite{journalversion}, the convergence condition of Gaussian BP for the linear Gaussian model is given as following.
\begin{mytheorem} \label{meanvector}
	The vector sequence
	$\left\{\textbf{v}^{\left(\ell\right)}\right\}_{l=0,1,\ldots}$ defined by (\ref{meanvectorupdate}) converges to a unique value
	for any initial value $\left\{\textbf{v}^{\left(0\right)}\right\}$ and initial covariance matrix inverse $\left[\textbf{C}_{f_{i,j}\to i}^{\left(0\right)}\right]^{-1}\succeq \mathbf 0$ if and only if $\rho\left(\mathbf {Q}\right)<1$.
\end{mytheorem}
The above condition subsumes the walk-summability condition as shown in   \cite{journalversion}.
However, checking this condition requires computation of the spectral radius of $\textbf Q$ with a computational complexity of $\mathcal{O}((\sum_{i=1}^{|\mathcal{V}|}N_i)^3)$, which conflicts with the low computation complexity purpose for  distributed inference.

We next  present a sufficient convergence condition and demonstrate a distributed way to verify this condition.
Since $\textbf{Q}$ is a square matrix, we have
$\rho\left(\textbf{Q}\right)\leq
\sqrt{\rho(\textbf{Q}\textbf{Q}^T)}$, and therefore $\rho(\textbf{Q}\textbf{Q}^T)<1$ is a sufficient condition for the convergence of
$\textbf v^{\left(\ell\right)}$.

According to the construction of    $\textbf{Q}_{j}$, which is
a row block matrix with block components  $\textbf 0$ and
$\textbf{C}^{\ast}_{j\to f_{i,j}}$ where
$f_{k,j}\in\B\left(j\right)\setminus f_{i,j}$, we have $$\textbf{Q}_j\textbf{Q}_i^T=\mathbf 0, \quad
\forall i\neq j.$$
Therefore, $\textbf{Q}\textbf{Q}^T$ is a block diagonal matrix with block diagonal elements $\textbf{Q}_{j}\textbf{Q}_{j}^T$, $j=1,\ldots, M$.
Following the analysis above (\ref{v2fm4}), $\textbf Q_j$ is a matrix with entries $\textbf{C}^{\ast}_{j\to f_{i,j}}$ or $\textbf 0$, therefore $\rho(\mathbf {Q}_j\mathbf {Q}_j^T)$ can be computed locally at node $j$.

From Theorem 1, by setting
$[\textbf{C}^{(0)}_{f_{i,j}\to i}]^{-1}\succeq \textbf{0}$, $[\textbf{C}^{(l)}_{f_{i,j}\to i}]^{-1}$ converges to a unique positive definite matrix.
Then, according to (\ref{v2fV}),
$\textbf{C}^{(\ell)}_{j\to f_{i,j}}$ converges to a unique positive definite matrix.
Once $\rho(\textbf{Q}\textbf{Q}^T)<1$ for all $i\neq j$, we know $\textbf{v}^{(\ell)}_{j\to f_{i,j}}$, which is an entry in $\textbf v^{\left(\ell\right)}$, converges.
Then, we can conclude $\bm \mu_i^{(l)} $ in (\ref{beliefu}) converges to a fixed point $\bm \mu_i^{\ast} $.
It has already been shown that, once Gaussian belief propagation converges, it converges to the centralized optimal estimate  \cite{journalversion, WalkSum1} for both the linear Gaussian model and the GMRFs.

The above analysis also applies to Gaussian BP on GMRFs.
We then have the following two theorems for the linear Gaussian model and GMRFs, respectively.
\begin{mytheorem} \label{meanvector}
	For the linear Gaussian model, the marginal mean computed by Gaussian BP converges to the exact mean
	if $\rho\left(\mathbf {Q}_j\mathbf {Q}_j^T\right)<1$ and  $[\textbf{C}^{(0)}_{f_{i,j}\to i}]^{-1}\succeq \textbf{0}$ for all $j\in \mathcal{V}, i\in \mathcal B(j)$.
\end{mytheorem}

\begin{mytheorem} \label{meanvector}
	For the GMRF, the marginal mean computed by Gaussian BP converges to the exact mean
	if $\rho\left(\widetilde{\mathbf {Q}}_j
	\widetilde{\mathbf {Q}}_j^T\right)<1$ and  $\Delta J^{(\ell)}_{j\to f_{i,j}}= {0}$ for all $j\in \mathcal{V}, i\in \mathcal B(j)$ .
\end{mytheorem}

\section{Conclusion}\label{conclusion}
This paper has established  distributed sufficient  convergence conditions for  Gaussian belief propagation  for both the linear Gaussian model and Gaussian Markov random fields.
Compared with existing convergence conditions for Gaussion belief propagation that need to gather  information for centralized computation, the proposed distributed sufficient condition only involves local computation at each node, which fits well  distributed computation.

\section{Acknowledgment} 
{We would like to acknowledge support for this project
	from the National Science Foundation (NSF grant  CCF-1513936). }

\end{document}